\documentstyle[11pt,aaspp4]{article}

\slugcomment{To appear in  The Astrophysical Journal}
\lefthead{Rodri\'{\i}guez-Ardila et al.}
\righthead{The Narrow Line Region of Narrow-Line Seyfert 1 Galaxies}
\newcommand{\lb}{$\lambda$}
\newcommand{\kms}{km s$^{-1}$}
\newcommand{\feii}{\ion{Fe}{2}}

\newcommand{\cm}{cm$^{-3}$}
\newcommand{\hbeta}{H$\beta$}
\newcommand{\halfa}{H$\alpha$}
\newcommand{\ii}{\'{\i}}

\begin{document}

\title{The Narrow Line Region of Narrow-Line Seyfert 1 Galaxies\altaffilmark{1}}

\author{A. Rodr\'{\i}guez-Ardila\altaffilmark{2,3}} 
\affil{Departamento de Astronomia -- UFRGS. CP 15051, Porto Alegre, Brazil}
 
\author{Luc Binette}
\affil{Instituto de Astronom\'{\i}a,  UNAM, Ap. 70-264, 04510 D.F., Mexico}

\author{Miriani G. Pastoriza}
\affil{Departamento de Astronomia -- UFRGS. CP 15051, Porto Alegre, Brazil}

\and

\author{Carlos J. Donzelli\altaffilmark{2}}
\affil{IATE, Observatorio Astron\'omico, Universidad Nacional de C\'ordoba\\
Laprida 854, 5000, C\'ordoba, Argentina}

\altaffiltext{1}{Based on observations made at CASLEO. Complejo
Astron\'omico El Leoncito (CASLEO) is operated under agreement
between the Consejo Nacional de Investigaciones Cient\'{\i}ficas
y t\'ecnicas de la Rep\'ublica Argentina and the National
Universities of La Plata, C\'ordoba and San Ju\'an.}
\altaffiltext{2}{Visiting Astronomer at CASLEO Observatory} 
\altaffiltext{3}{CNPq Fellow}

\begin{abstract}
This work studies the optical emission line properties and physical 
conditions of the narrow line region (NLR) of seven narrow-line 
Seyfert 1 galaxies (NLS1) for which high signal-to-noise spectroscopic
observations were available. The resolution is 340 \kms\ (at \halfa) over 
the wavelength interval 3700 -- 9500\,\AA, enabling us to separate
the broad and narrow components of the permitted emission lines.
Our results show that the flux carried out by the narrow component of 
\hbeta\ is, on average, 50\% of the total line flux. As a result, the 
[OIII] \lb5007/\hbeta\ ratio emitted in the NLR varies from 1 to 5, instead 
of the universally adopted value of 10. This has strong implications for the 
required spectral energy distribution that ionizes the NLR gas. 

Photoionization models that consider a NLR composed of a combination
of matter-bounded and ionization-bounded clouds are successful at
explaining the low [OIII] \lb5007/\hbeta\ ratio and the weakness
of low-ionization lines of NLS1s.  Variation of the relative
proportion of these two type of clouds nicely reproduce the dispersion
of narrow line ratios found among the NLS1 sample.  Assuming similar
physical model parameters of both NLS1s and the normal Seyfert 1
galaxy NGC\,5548, we show that the observed differences of emission
line ratios between these two groups of galaxies can be explained,
to a first approximation, in terms of the shape of the input
ionizing continuum.  Narrow emission line ratios of NLS1s are better
reproduced by a steep power-law continuum in the EUV -- soft X-ray
region, with spectral index $\alpha \sim -2$.  Flatter spectral
indices ($\alpha \sim -1.5$) match the observed line ratios of
NGC\,5548 but are unable to provide a good match to the NLS1
ratios. This result is consistent with ROSAT observations of NLS1s,
which show that these objects are characterized by steeper power-law
indices than those of Sy1 galaxies with strong broad optical lines.
\end{abstract}

\keywords{galaxies: Seyfert  -- galaxies: nuclei -- X-rays: galaxies}

\section{Introduction}
Narrow-Line Seyfert 1 Galaxies (hereafter NLS1) are a peculiar group of 
AGNs where the permitted optical lines show full width half-maximum 
(FWHM)  not exceeding 2000 \kms,
the [\ion{O}{3}] \lb5007/\hbeta\ ratio is $<$ 3 and the UV-VIS spectrum is
usually very rich in high ionization lines and \feii\ emission multiplets. 
In the soft X-ray band, NLS1s have generally much steeper continuum slopes
and rapid variability (Boller, Brandt \& Fink 1996, hereafter BBF96). 
Recently, Leighly (1999) found that the hard X-ray photon index is 
significantly steeper in NLS1s compared with that of normal Seyfert 1s, 
and that soft excess emission appears considerably more frequently in 
NLS1s than in Seyfert 1 (hereafter Sy1) galaxies with broad optical lines.

It is not known at present the origin of the narrowness of broad
permitted lines in NLS1s. Osterbrock \& Pogge (1985); Ulvestad,
Antonucchi \& Goodrich (1985) and Stephens (1989) suggest that if the
velocities in the BLR of Seyfert 1s were largely confined to a plane,
the NLS1 galaxies could be understood as cases in which the line of
sight is nearly perpendicular to this plane. BBF96 state, on the
other hand, that if the gravitational force from the central black
hole is the dominant cause of the motions of Seyfert BLR clouds,
narrower optical emission lines will result from smaller black hole
masses provided the characteristics BLR distance from the central
source does not change strongly with black hole masses.

However, Rodr\'{\i}guez-Pascual, Mass-Hesse \& Santos-Ll\'eo (1997)
report the detection in NLS1 galaxies of broad components with FWHM
around 5000 \kms\ for the high ionization UV permitted lines such
as Ly$\alpha$,
\ion{C}{4} \lb1550 and \ion{He}{2} \lb1640. This result indicates that gas 
moving at velocities comparable to those found in typical Sy1 galaxies does 
indeed exist in NLS1s. In the optical region, they found ``broad'' 
components with FWHM less than 3000 \kms, narrower than the broadest UV 
component in the same objects. Nonetheless, deblending the optical permitted 
lines in NLS1s is difficult because no transition between the 
narrow and broad components is observed. This shortcoming has strong 
influences in, for example, the analysis of the narrow line region (NLR) 
due to the large uncertainties in determining the fraction of \halfa\ 
and \hbeta\  which originates from low ionization material. 

Up to now, most studies of the NLR in NLS1s assume that the flux emitted 
by the narrow \hbeta\ equals  10\% of the flux of [\ion{O}{3}] \lb5007 
(Osterbrock \& Pogge 1985; Leighly 1999). This assumption is based on the 
results obtained from Seyfert 2 and intermediate Sy1 galaxies (e.g. Koski 
1978; Cohen 1983). But in recent years, growing observational evidence 
points out to the existence of differences between the NLR of normal Sy1 
and Sy2 galaxies (Schmitt \& Kinney 1996; Schmitt 1998), making the 
above assumption highly uncertain. In addition, fixing the [\ion{O}{3}] 
\lb5007/\hbeta\ ratio to 10 implies ignoring the large scatter in the 
value of this ratio observed in normal Sy1s (2 to 19, see for example 
Rodr\'{\i}guez-Ardila, Pastoriza \& Donzelli 1999, hereafter Paper\,I) 
and overlook the influences that this ratio could have in the energetics 
and physical conditions of the NLR of these objects. 

Due to the above reason, the main purpose of this paper is to seek
additional constrains in order to estimate the actual contribution of
the narrow \hbeta\ flux to the total \hbeta\ emission line and study
the implications that the newly adopted values could have on narrow
line ratios and the physics of the NLR of NLS1s.

The present work is organized as follows. In Section~\ref{observ} we
describe the sample of NLS1s used in this paper. Section~\ref{gauss}
presents the decomposition into narrow and broad components carried
out in the Balmer lines of the NLS1 galaxies. Photoionization models
that successfully reproduce the observed line ratios of NLS1s are
presented in Section~\ref{MBIB}. A discussion of the main results
appears in Section~\ref{bla} and the conclusions are presented
in Section~\ref{fin}.

\section{Observations}\label{observ}

Long-slit spectroscopic observations of seven NLS1 galaxies covering
the spectral region 3700 \AA\ - 9500 \AA\ were obtained with the 2.15
m telescope of the Complejo Astronomico El Leoncito (CASLEO) using a
TEK 1024 $\times$ 1024 CCD detector and a REOSC spectrograph. Two
gratings of 300 l/mm with blaze angles near 5500 \AA\ and 8000 \AA,
respectively, were used in order to fully cover the spectral
interval 3700 -- 9500 \AA. The spatial scale of this setup is
0.95\arcsec/pix, with an instrumental resolution of 7 \AA\ FWHM.
A slit width of 2\arcsec.5 oriented in the East-West
direction and crossing the center of the galaxies was employed.  The
galaxies and standard stars were observed near the zenith (air masses
$<$ 1.2). A complete log of the observations and reduction procedure
are described in Paper\,I.

\begin{figure}
\plotone{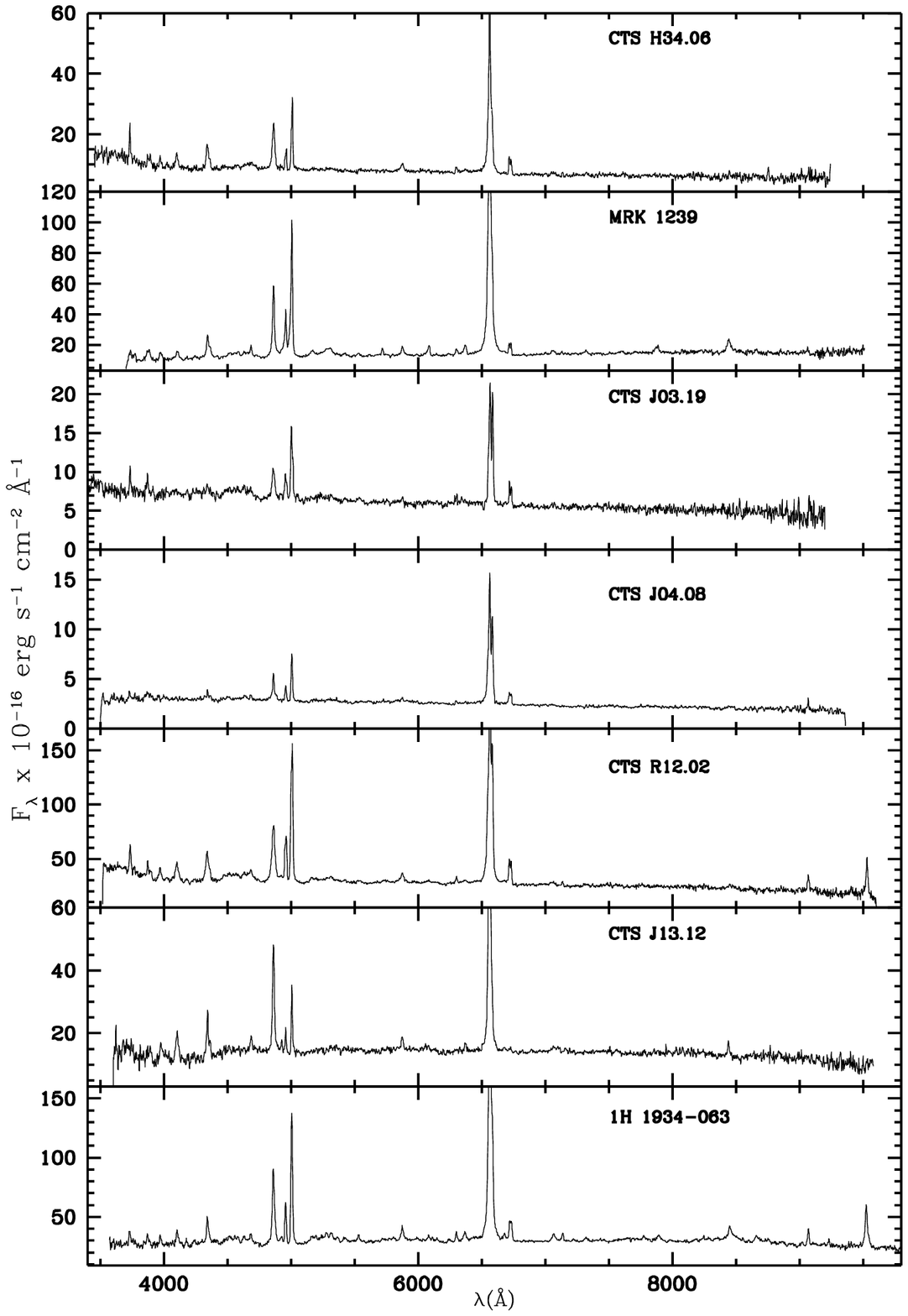}
\caption{Sample of NLS1 galaxy spectra used in this work. \label{espectros}}
\end{figure}

CTS\,J03.19, CTS\,J04.08, CTS\,J13.12
and CTS\,H34.06 were classified as NLS1s by us based on the appearance
of their respective optical spectrum. The criterion used in
this classification is that the permitted lines be slightly broader
than the forbidden lines, following the definition of Osterbrock \&
Pogge (1985). MRK\,1239, CTS\,R12.02 (=NGC\,4748) and 1H\,1934-063 are 
well known NLS1s objects. For this reason, we consider that our sample 
is optically selected, in contrast to most NLS1s published in the 
literature, which are based on X-ray selected objects. The luminosities 
of the galaxies are low to intermediate and their radial velocities are 
not larger than 16,000 \kms. Figure~\ref{espectros} shows the spectra of 
the seven galaxies of the current study, already corrected for redshift.

\begin{figure}
\plotone{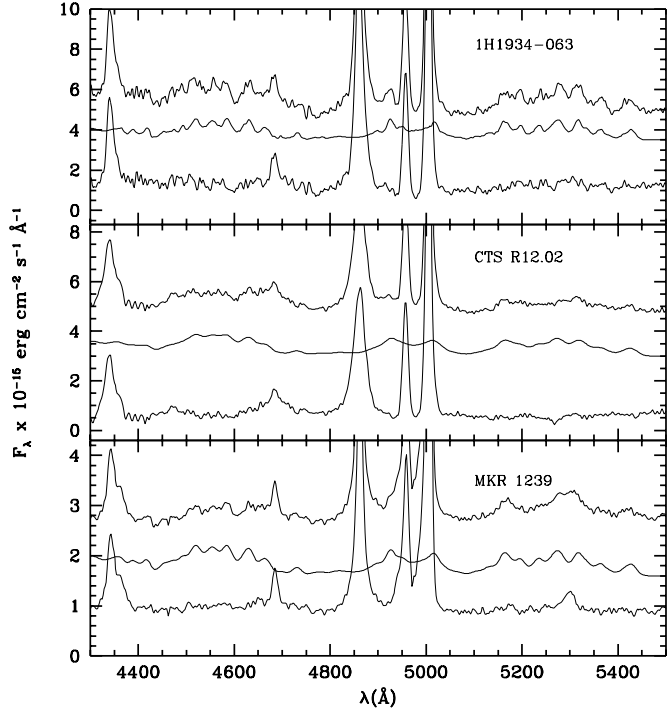}
\caption{Example of the \feii\ subtraction procedure carried
out to the NLS1 galaxies 1H\,1934-063 (top panel), CTS\,R12.02 (middle
panel) and MRK\,1239 (lower panel). In each panel from top to bottom
are the observed spectrum with the \feii\ emission, the \feii\ template 
that best matches this emission and the residual Seyfert spectrum after 
subtracting the \feii\ contribution. The spectra have been displaced by 
a constant factor for visualization purposes. In the residual spectrum 
note the flat continuum at both sides of H$\beta$ and the appereance of 
the He\,II \lb4686 line not clearly seen in the contaminated spectrum. 
\label{iron_remov}}
\end{figure}

\feii\ emission, which is particularly strong in NLS1s, may alter the 
flux and width of \hbeta\ + [\ion{O}{3}] \lb\lb4959,5007. For this 
reason we have carefully removed the \feii\ multiplets following the 
method described in Boroson \& Green (1992). It consists of constructing a 
\feii\ template by removing the lines which are not of \feii\  from
the spectrum of IZw1, a NLSy1 galaxy widely known for the strength
of the \feii\ emission and the narrowness of its ``broad lines.'' 
For this purpose, a high signal to noise spectrum of IZw1 covering 
the spectral range 3500-6800 \AA\ was taken in one of the observing runs. 
After isolating the \feii emission, the template was broadened by convolving 
it with a Gaussian profile having a FWHM similar to that of
the H$\beta$ line and scaled to match the observed \feii\ emission of the
corresponding galaxy. Figure~\ref{iron_remov} shows the above procedure for 
1H\,1934-063 (upper panel), CTS\,R12.02 (middle panel) and MRK\,1239
(lower panel). In each case it is shown the observed spectrum, the 
\feii\ template and the resulting spectrum after removing the \feii\
emission. It can be seen that the most important \feii\ features
at both sides of \hbeta\ have been cleanly removed leaving as a residual
the true \ion{He}{2} \lb4686 line profile and, in MRK\,1239, the
[FeIXV] \lb5302 line.

\section{The Gaussian Description of The Emission Line Profiles}\label{gauss}

\subsection{Results of line profile fitting}\label{fit}

In order to characterize the emission line profiles of the NLS1s we 
have assumed that they can be represented by a single or a 
combination of Gaussian profiles. The {\sc liner} routine (Pogge \& Owen 
1993) which is a $\chi^{2}$ minimization algorithm that fits as many as eight 
Gaussians to a line profile, was used for this purpose.

\begin{figure}
\plotone{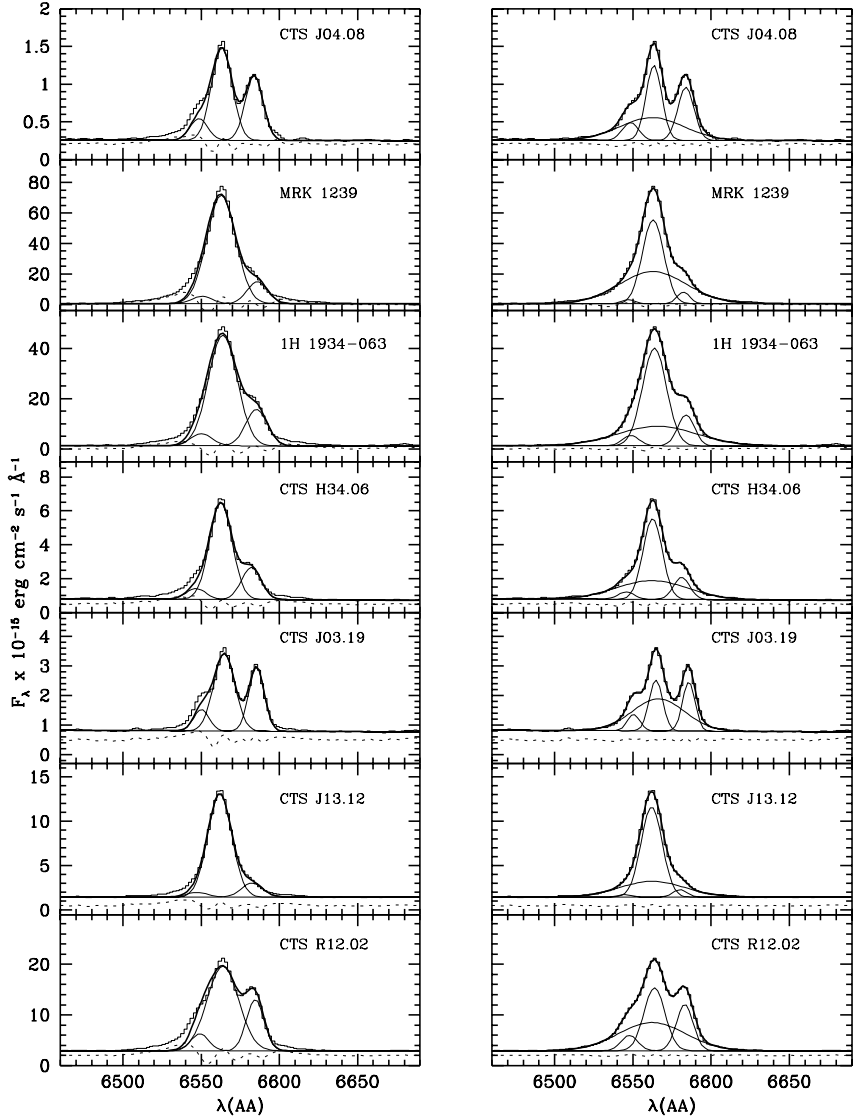}
\caption{Gaussian fitting to the \halfa\ + [\ion{N}{2}]
\lb\lb6548,6584 line profiles for the seven NLS1s galaxies of the sample.
The left panels show \halfa\ fitted with a single component. The
addition of a second, broader component to this line enhances the
quality of the fit, as is shown in the right panels. The histogram is
the data, the thin lines represent the individual components and the
underlying continuum, the dashed line represents the residuals of
the fit and the thicker line the synthetically calculated profile. \label{ha_fit}}
\end{figure}

As a first steep we tried to fit the \halfa\ emission line with a
single Gaussian component. \halfa\ is the strongest optical permited
line and is located in a spectral region with the highest
S/N. It is therefore the best place to look for broad components
to the permitted optical lines of NLS1s. The best fit obtained is
shown in the left panels of Figure~\ref{ha_fit}. The thick line
represents the synthetically calculated profile and the dotted
line the residuals of the fit.  It is clear that this simplest
representation cannot adjust adequately the wings of \halfa\
although its core is nicely fitted in most of the objects. We then
tried to adjust a second Gaussian to \halfa. The best solution found
are shown in the right panels of the same figure. Now the residuals of
the fit are quite similar to the noise level around
\halfa, implying that a narrow (FWHM $\sim$ 600 \kms) plus a broad 
component (FWHM $\sim$ 2500 \kms) give a convincingly better description of the 
observed profiles. In the above fits the only constraints applied were 
that [\ion{N}{2}] \lb6548 and [\ion{N}{2}] \lb6584 be of equal FWHM and 
that their flux ratio and wavelength separation be equal to their 
theoretical value (1:3 and 36 \AA, respectively). 

\begin{figure}
\plotone{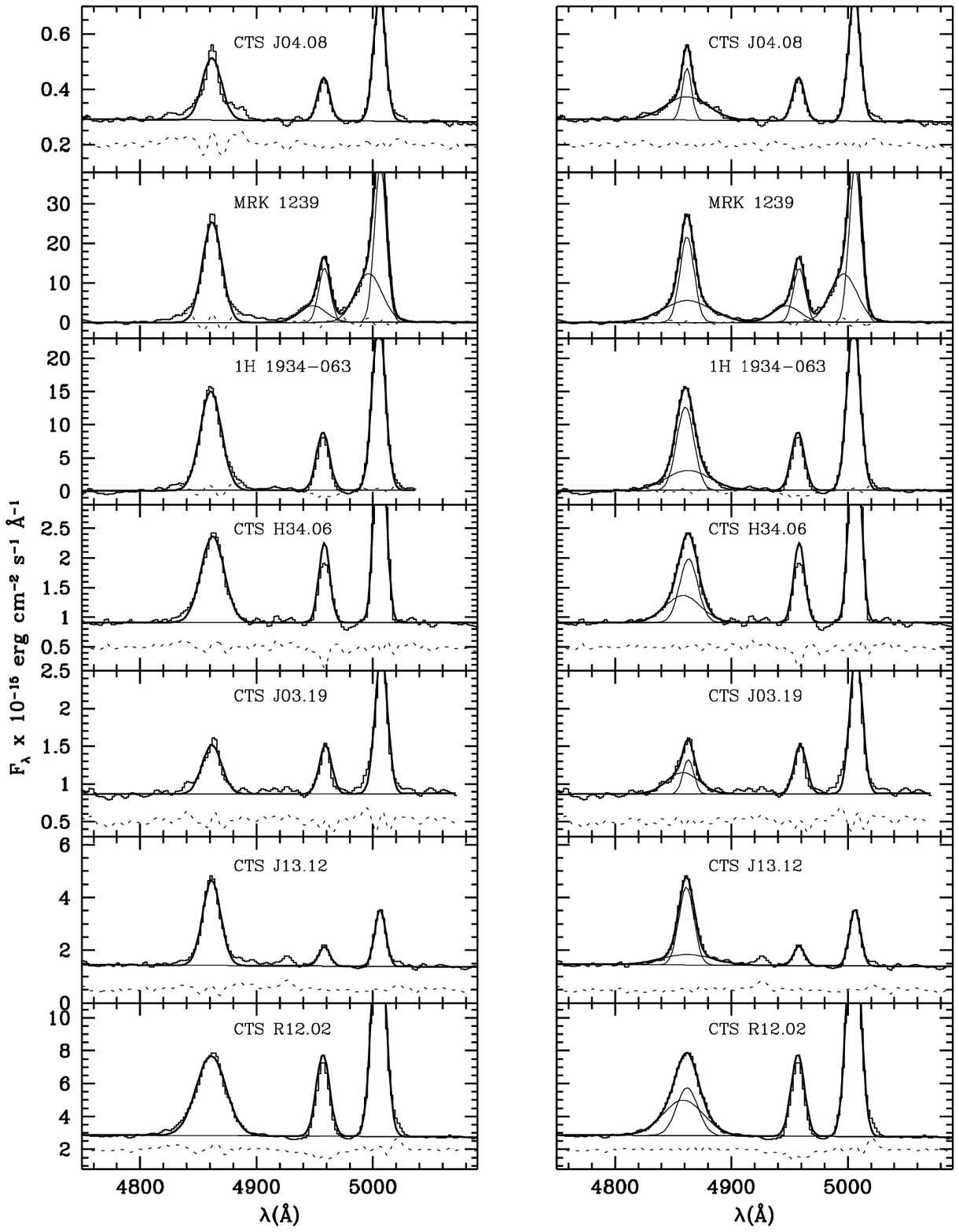}
\caption{The same as Figure~\ref{ha_fit} but to the
\hbeta\ + [\ion{O}{3}] \lb\lb4959,5007 lines. \label{hb_fit}}
\end{figure}

The same decomposition was applied to \hbeta\ + [\ion{O}{3}]
\lb4959,5007.  As for \halfa, the fit of \hbeta\ with a single
Gaussian gives a poor representation of this line and it was
necessary to include an additional component to represent adequately
the observed profile. Figure~\ref{hb_fit} shows the results of this
decomposition. The residuals (dashed line) of the dual component fit
(right panels) are significantly improved with respect to the fit with
a single Gaussian (left panels).

Columns 2 to 7 of Table~\ref{fwhm} list the FWHM (in \kms) of the
emission lines measured from the Gaussian fitting. These values were
obtained from the subtraction, in quadrature, of the observed FWHM and
that of the instrumental profile, measured from the comparison lamp
lines (FWHM $\sim$ 360 \kms\ at \halfa). The flux ratio of the
narrow to the broad component of \hbeta\ for each galaxy is listed in
Column 8 and the [\ion{O}{3}] \lb5007 flux, relative to the narrow
\hbeta\ component, is in Column 9. In all but one case (MRK\,1239),
the center of the narrow components was coincident with the systemic
velocity of the corresponding galaxy. MRK\,1239 presents a second
blueshifted (with an outflow velocity of 600 \kms\ with respect to the
nucleus) broader (FWHM $\sim$ 1580 \kms) component that is also present
in \halfa\ and \hbeta\ (see Figures~\ref{ha_fit} and~\ref{hb_fit})
and that we associated to the NLR .

\begin{figure}
\plotone{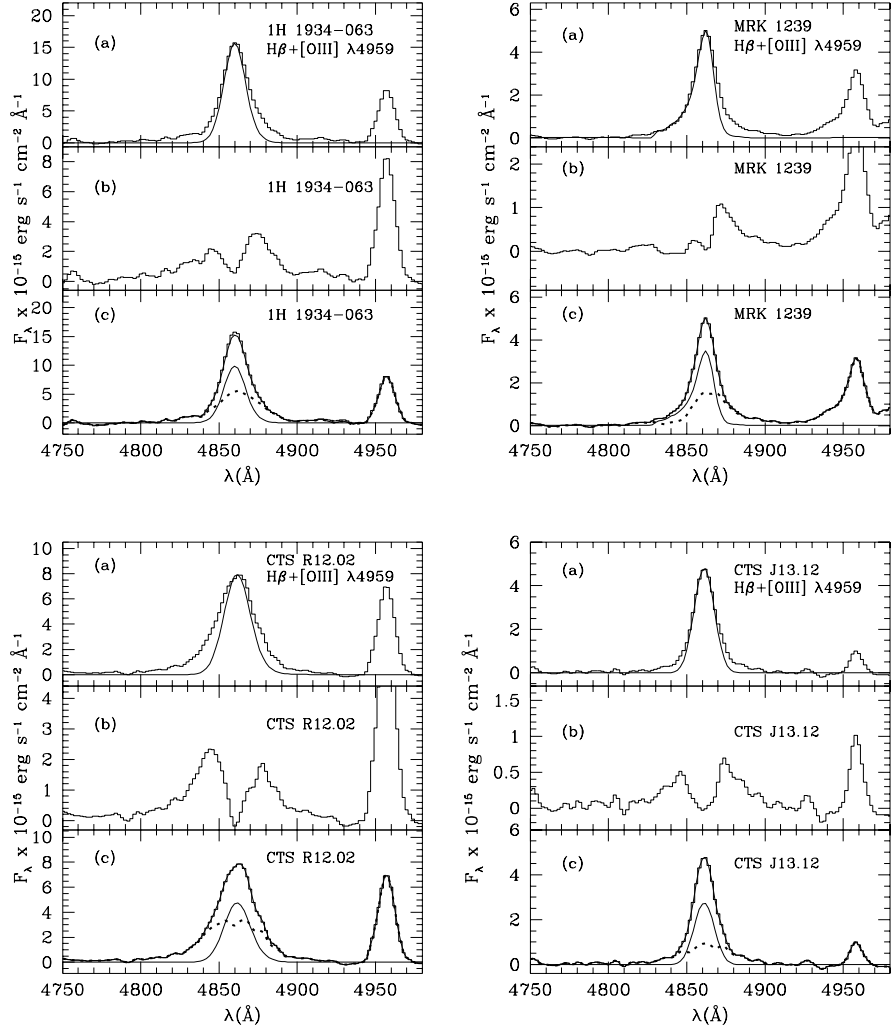}
\caption{Example of the deblending procedure applied to the
\hbeta\ line using [\ion{O}{3}] \lb5007 as a template for the narrow
components. For each galaxy, panel (a) shows the template narrow line
scaled to the peak intensity of the observed \hbeta\ (histogram); panel
(b) shows the residuals after subtracting the template line. They consist of
a broad wing, attributed to the presence of a second broader component of 
\hbeta\ and a strong absorption due to an overestimation of the contribution 
of the narrow component. Reducing the strenght of the narrow component
so that no absorption is observed in the residuals (panel c) leaves a
pure broad component (dashed line) very similar to that found using
the Gaussian decomposition. \label{test_profiles}}
\end{figure}

It is important to stress that the same result, --i.e a narrow component
plus a broad one in the permitted lines of \halfa\ and \hbeta\ --  would be 
obtained if instead of using the Gaussian decomposition technique we had 
employed the [\ion{O}{3}] \lb5007 line as a template representative of the 
NLR profiles. This test was applied to the \hbeta\ line of the galaxy sample
in order to see if the broad component found in the permitted lines
using the Gaussian representation were not an artifact of the fitting
procedure. For this purpose the [\ion{O}{3}] \lb5007 line was isolated and
normalized to the peak intensity of the \hbeta\ line to represent
the maximum allowed \hbeta\ contribution from the NLR to the observed profile.
When subtracted, the residuals consist basically of a broad wing and a
strong absorption coincident with the peak position of the template profile,
such as shown in Figure~\ref{test_profiles} for the NLS1 galaxies
1H\,1934-063, CTS\,R12.02, CTS\,J13.12 and MRK\,1239.

We interpret the broad wings as a clear evidence of the presence of a
broad component similar to that observed in normal Seyfert 1 galaxies but
with a smaller FWHM, supportting the results of Rodr\ii guez-Pascual, 
Mass-Hesse and Santos-L\'eo (1997); Gon\c calves, V\'eron \& V\'eron-Cetty 
(1999) and Nagao et\,al. (1999) who also report a broad component in the
permitted lines of NLS1s. The absorption is interpreted as due to an overestimation 
of the NLR component. Scaling the template profile in order to eliminate the 
absorption in the residuals leaves a pure broad component very similar in 
intensity and width as that found in the Gaussian decomposition (dashed line
of Figure~\ref{test_profiles}). In fact, the values of FWHM and line fluxes 
obtained using this approach is essentially the same as those obtained formely. 

We conclude from the above test that the broad feature observed in 
\hbeta\ and \halfa\ is real and it is interpreted here as the
contribution from the BLR to the \hbeta\ line.
 
\begin{figure}
\plotone{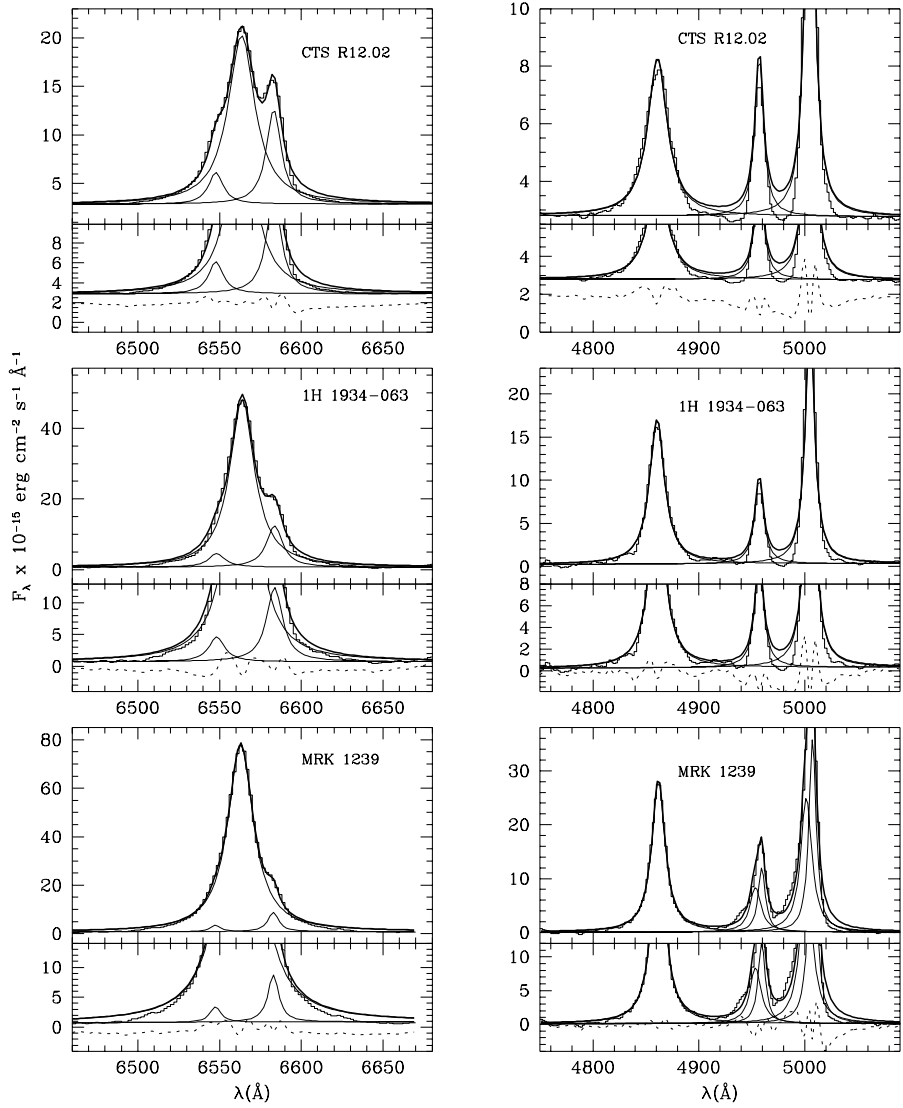}
\caption{Fitting of  \halfa\ + [\ion{N}{2}] (left panels) and \hbeta\ + 
[\ion{O}{3}] \lb\lb4959,5007 (right panels) for CTS\,R12.02 (upper panel), 1H\,1934-063 
(middle panel) and MRK\,1239 (lower panel) with Lorentzian profiles. The histogram is
the observed data, the thin lines represent the individual components and the
underlying continuum, the thick line the resultant synthetic profile and the dashed line 
represents the residuals of the fit. Note that in all cases the wings of the
Lorentzians are more extended than the observed profiles. 
\label{lorentz}}
\end{figure}

Some authors (Moran, Halpern \& Helfand 1996, Gon\c calves, Veron-Cetty 
\& Veron 1999) have suggested that NLS1s have more nearly Lorentzian, 
rather than Gaussian, profiles, as evidenced by their cusped peaks and 
broad wings, mainly in \hbeta\ and [\ion{O}{3}] \lb5007 lines and even in 
\halfa. We have tested this hypothesis by fitting one-component 
Lorentzian profiles to the Balmer and adjacent lines. The best
solution, plotted in Figure~\ref{lorentz} for CTS\,R12.02,
1H\,1934-063 and MRK\,1239 shows that this description does not
provide a satisfactory fit to the emission lines. Very similar results
were obtained for the other NLS1s of our sample. In all cases, the
wings of the Lorentzians are more extended than the wings of the
observed profiles. This effect is more pronounced in \halfa\ (left
panels of Figure~\ref{lorentz}) than in \hbeta\ (right panels), giving
the apparent impression that this latter line is better fitted by a
Lorentzian profile. We attribute this effect to the lower S/N of the
\hbeta\ region, which reduces the contrast of faint broad component
wings relative to the narrow line. The poor fit  obtained for the [\ion{O}{3}]
\lb\lb4959,5007 lines demonstrate that the Lorentzian profiles 
are unsuitable for representing the NLR profiles of NLS1s.

At this point it is important to note that the choice of the Gaussian 
profile as representative of the form of the observed emission lines was
due to its simplicity and lack of physical reasons to adopt another 
particular form. We must be aware that the decomposition of the permitted 
lines into narrow and broad components, particularly in NLS1s, is a very
uncertain task. As Evans (1988) noted, the fitting of symmetrical functional 
forms to the observed profiles may lead to the appereance of ``components'' in the
center and wings of the lines but that, in certain circumstances, cannot 
bear any physical meaning. In addition, the observed profile can be
the result of the superposition of many emitting regions  located
along the line of sight, each of them with a different intrinsic
profile. For these reasons, one has to look with caution the
deblending into broad and narrow components obtained for the permitted
lines. Nonetheless, the fact of having obtained similar results using two
independent methods give additional support to our interpretation.

\subsection{Meaning of the Narrow and Broad Components in NLS1s}\label{nabc}

In the previous section we showed that the permitted lines of NLS1s
can be decomposed into a narrow and a broad component, as is usually
carried out in normal and intermediate Seyfert 1 galaxies. But what is
the real meaning of this? A Gaussian representation as such, is only a
mathematical way of fitting approximately the data. Additional
evidence and a rationale are needed in order to identify each
Gaussian component with physically different regions. Also, we must be
aware that the number of Gaussian components adjusted to a given line
basically depends on the spectral resolution and the S/N within
the region of the fit.

Two important points call our attention regarding the values presented in 
Table~\ref{fwhm}. First, the FWHM of the ``broad'' components in \halfa\ 
and \hbeta\ are rather similar within the same galaxy and  from object
to object. Second, the FWHM of the narrow component of permited lines is
similar to that of the forbidden ([\ion{O}{3}] and [\ion{N}{2}])
lines. They are also larger than the width of the instrumental profile
(FWHM $\sim$ 360 \kms).

It is generally accepted that the kinematics of the BLR clouds is
largely dominated by the gravitational potential of the central black
hole. The FWHM of the lines emitted in that region is a measure of the
dispersion of velocities of the emitting gas, along the observer's
line of sight, which represents also the depth of the gravitational
potential well in which the emitting clouds find themselves. For
this reason, it is expected that lines formed in the same spatial
region will have similar FWHM and emission line profiles.

A similar argument can be applied to the narrow lines. However, the NLR is
much more extended and located farther out from the central source (1--100 
pc) so these clouds are immersed in a shallower gravitational
potential dominated by the galaxy bulge.  

Therefore, the consistency of the width and profile form of the broad
components of \halfa\ and \hbeta\ and the similarity in width of
the narrow permitted and forbidden lines, within the same galaxy,
allow us to associate the narrow and broad Gaussian components of the
permitted lines to the integrated line emission from the NLR and BLR,
respectively.

A comparison of the FWHM of the narrow lines of our sample with those
measured with a similar setup in normal Seyfert 1s (Cohen 1983;
Stephens 1989; Puchnarewicz et\,al. 1997) shows that no
differences seem to exist in the NLR kinematics between these two
groups of galaxies.

The main difference lies in the BLR. The FWHM of the broad component
is not only significatively smaller than that of the normal Sy1s but
also its relative contribution to the total flux of the line is
greatly reduced in these objects.  In Column 8 of Table~\ref{fwhm} we
have listed the ratio of the flux associated to the narrow component
to that of the broad one. It can be seen that on average, this
ratio is very near unity, meaning that 50\% of the total line flux
is due to the narrow component. For comparison, this ratio is around 0.1
in typical Sy1 galaxies. 

The above results by themselves do not represent a major departure
from our current picture of NLS1s. However, a significant difference
emerges when we consider line ratios between the narrow component of
\hbeta\ and [\ion{O}{3}] \lb5007.  In effect, since Osterbrock \& 
Pogge (1995) and up to very recently (Leighly 1999), it has been 
assumed that the contribution in flux of the narrow component of 
\hbeta\ to the NLR spectrum of a given object equals 10\% of the 
flux of [\ion{O}{3}] \lb5007. This assumption is based on the fact 
that [\ion{O}{3}] \lb5007/\hbeta\ is, on average, $\sim$ 10 in Seyfert 
2 galaxies (Veilleux and Osterbrock 1987). Nonetheless, growing 
evidence of important differences between the NLR of Sy1 and Sy2 
galaxies have appeared in the literature (Schmitt \& Kinney 1996, 
Schmitt 1998; Paper I). In addition, NLS1s are recognized as a 
subclass within the realm of Seyfert 1s due to their peculiar 
properties, making very unlikely that the above assumption really 
holds in these objects, as is shown in column 9 of Table~\ref{fwhm}, 
which lists the [\ion{O}{3}] \lb5007/\hbeta(narrow) ratio found from 
our decomposition of line profiles. These values also show that there 
is a wide intrinsic dispersion in the [\ion{O}{3}] \lb5007/\hbeta\ 
ratio of NLS1s, ranging from 0.8 to 5.

Since the new values of the narrow [\ion{O}{3}] \lb5007/\hbeta\ ratio
are drawn from the decomposition into narrow and broad components of the
Balmer lines, it is important to discuss the uncertainties inherent
to the deblending process. Strictly speaking, one should be inclined to
treat the flux associated to the narrow component of the permitted 
lines as a lower limit, being the actual flux between this value and that
obtained from the total flux of the line. The latter option would be the 
case if no contribution from the BLR were present, as was initially suggested 
by Osterbrock \& Pogge (1985) in order to explain the absence of broad 
permitted lines in NLS1s. Under the last circumstance, the resulting 
[\ion{O}{3}] \lb5007/\hbeta\ ratio would be even smaller but would not 
change drastically. It would now fall into the interval 0.5 -- 1.7, 
making the departure from the ratios found in normal Seyfert 1 galaxies 
even stronger. 

One can argue that the deblending of the permitted lines 
overestimated the flux of the narrow \hbeta\ component, making the
[\ion{O}{3}] \lb5007/\hbeta\ ratio appear smaller that it really does.
That is, instead of being a lower limit, it represents an upper limit. 
Under this circumstance, the actual flux of the narrow \hbeta\ would range 
between this upper limit and a small fraction of the total line flux. 
Physically, this is highly improbable, at least for two reasons. First,
\hbeta\ is emitted by every gas component, so the narrow \hbeta\ cannot
be narrower than the narrowest NLR line. This is in accord to the 
values of Table~\ref{fwhm}, where the FWHM of \hbeta\ is similar to that
of [\ion{O}{3}] \lb5007 and of the same order of [\ion{N}{2}] \lb6584. 

Second, lets suppose that the actual
narrow \hbeta\ flux corresponds to 50\% of value obtained from the deblending
process. With this in mind,  the [\ion{O}{3}] \lb5007/\hbeta\ ratio would 
now fall in the interval 1.6 -- 10. But now a new problem will arise: the 
reddening measured from the Balmer decrement would be increased up to 0.7 mag. 
Such an increase in the E(B-V) would have a strong influence in the 
resulting emission line spectrum. Probably new physical 
processes and ionization mechanisms should need to be invoked in order to explain 
the observed spectrum. In addition it would not be clear in which emitting 
region the remaining ``narrow'' flux would be produced. In the BLR? in the
intermediate NLR? 

In the following sections, additional evidence supporting
the results obtained from the deblending process is given. Nonetheless,
it is clear that a large spatial resolution, such as that obtained with the HST,
is necessary in order to examine carefully the NLR of NLS1 galaxies. 


\section{Matter-Bounded and Ionization Bounded clouds in the NLR of NLS1s}\label{MBIB}

We have found that the NLR of NLS1s are
characterized by lower [\ion{O}{3}] \lb5007/\hbeta\ ratios, more
typical of some starburst or \ion{H}{2} galaxies than of canonical
Seyfert 1s or 2s. In addition, it was shown in Paper\,I that {\it i)}
NLS1s have intrinsically weak low ionization forbidden lines; {\it
ii)} the dominant mechanism of the NLR emission is photoionization
by a central source and {\it iii)} a wide range of densities
(10$^{3}$ \cm\ -- 10$^{6}$ \cm) must exist in the NLR of these
objects, with the larger densities associated to the inner NLR
regions where [\ion{O}{3}] and the high ionization lines are
emitted while the lowest values are associated  to the
[\ion{O}{1}] and [\ion{S}{2}] emitting regions.

A scenario which springs naturally from  the above results invokes
the presence of at least two types of clouds. A denser, inner set
of high excitation matter-bounded (MB) clouds are required to enhance 
the temperature sensitive [\ion{O}{3}] 4363\AA\ line relative to 5007\AA\ 
(due to collisional deexcitation) and also matter-bounded in order {\it
not} to emit low excitation lines (from low ionization species such as
\ion{O}{1}, \ion{S}{2}, \ion{N}{2}...). The MB component is where most
of the [\ion{O}{3}] emission and high ionization lines such as
\ion{He}{2}, [\ion{Ne}{5}], [\ion{Fe}{7}] and [\ion{Fe}{10}] would be
produced. To account for the lower excitation lines, it is necessary
to consider a second component consisting of outer, less dense set of
ionization-bounded (IB) clouds, responsible for the production of
low-ionization lines such as [\ion{O}{1}], [\ion{N}{2}] and
[\ion{S}{2}] (whose doublet ratio points to low densities) and some
[\ion{O}{3}] as well. These clouds must lie at a much larger distance
in order that the ionization parameter be much smaller than the MB
component as discussed in Rodr\'{\i}guez-Ardila, Pastoriza \& Maza (1998,
hereafter RAPM). Our starting hypothesis is that the
relative proportion of these two types of clouds combined with a
suitable choice of the input ionizing continuum will reproduce the
observed differences between the NLR of NLS1s and that of normal
Seyfert 1 galaxies.

This dual-component model have been successfully applied before to
account for the observed emission line ratios of the NLR and extended
narrow line region of many Seyfert 1 and 2 galaxies (Binette, Wilson
\& Storchi-Bergmann 1996, Binette et~al. 1997, RAPM, Aita-Fraquelli,
Storchi-Bergmann \& Binette 1999). Such a simple model is surely
an over-simplification relative to the wide range in parameters needed
to fully describe the NLR (to illustrate the complexity of the problem
see Moore \& Cohen 1996, Ferguson et\,al. 1997 and Komossa \& Schulz
1997). However, in the absence of sufficient observational constraints
on all possible models and geometries of the NLR, it appears to us to
be the simplest mean for taking into account the particular emission
line signature reported in RAPM, that is a high density high
excitation spectrum observed in conjunction with the presence of low
density low excitation lines.

\subsection{The Photoionization Models}\label{photmod}

In this section we test whether the simplified dual-component MB-IB
description can satisfactorily account for the anomalous line ratios
observed in NLS1s. We are particularly interested in determining  
which kind of spectral energy distribution (SED) observed in AGNs can 
best produce a [\ion{O}{3}] \lb5007/\hbeta\ ratio in the range 1 -- 5
and the strongest  optical emission lines, using 
similar physical parameters to those employed to model the NLR of normal 
Seyfert 1 galaxies. A detailed photoionization modeling of the NLS1 
emission line ratios is in preparation (Rodr\ii guez-Ardila, Binette \& 
Pastoriza 1999) 

Although there is no direct way to determine the intrinsic shape of
the ionizing continuum in the EUV domaine, recent work by Zheng
et\,al. (1997) and Laor et\,al. (1997) show that the ionizing
continuum, from the Lyman limit to soft X-ray energies, may be
characterized by a power-law of index $\alpha \sim -2$. This
result was derived from QSOs of intermediate redshifts but may
be extended to lower luminosity AGN such as Seyfert 1
galaxies. Korista, Ferland \& Baldwin (1997) suggest flatter indices
$\alpha \sim -1.5$, which have been used by Binette, Wilson \&
Storchi-Bergmann (1996) and RAPM, for instance, in their modeling of Seyfert 1
objects.  Meanwhile, ROSAT observations of NLS1s show that the soft
X-ray photon index is systematically steeper than that of Seyfert
1 galaxies with broad optical lines (BBF96; Foster \& Halpern 1996; 
Laor et al. 1997).

Due to the above reasons, we generated sequences of models
assuming a SED of broken power-laws of the form F$_\nu$ =
K$\nu^{\alpha}$ where:

\begin{equation}
\alpha=-1.4, ~13.6~eV \leq ~h\nu \leq ~1300 ~eV;  ~~\alpha=-0.4, ~h\nu \geq ~1300 ~eV 
\end{equation}
\begin{equation} 
\alpha=-2.2, ~13.6~eV \leq ~h\nu \leq ~2000 ~eV;  ~~\alpha=-1.1, ~h\nu \geq ~2000 ~eV
\end{equation}

Equation\,1 (SED\,1) corresponds to the fit made by Kraemer et\, al. (1998) to
the observed SED of the Seyfert 1 galaxy NGC\,5548. Equation\,2 (SED\,2) uses
the median values of the spectral indices found from ROSAT and ASCA data
for NLS1s (Leighly 1999).  

For each SED, the $A_{M/I}$ parameter, which characterizes
the relative proportion of MB and IB clouds, was varied from
0.04 to 11. Density estimates of the NLR of NLS1s based on density and
temperature line ratios (cf Paper\,I) show that MB clouds should have $n_{e}
\sim 10^{6}$ \cm\ while in IB clouds $n_{e} \sim 10^{3}$
\cm. The ionization parameter $U_{MB}$ at the illuminated face of the MB
clouds was initially set to 10$^{-1.5}$, following estimates by RAPM, 
Kraemer et\,al. (1998) and Kraemer et\,al. (1999) for normal Seyfert 1
galaxies.

The multipurpose code MAPPINGS\,Ic (Ferruit et\,al. 1997) was used to
compute the dual-component photoionization models.
Plane-parallel geometry was assumed given the relatively large
distance of the ionized clouds from the central source compared to the
geometrical depth of the clouds. The gas is assumed to be atomic and
the gas abundances solar. In order to let the inner set of clouds be
matter-bounded, a fraction F$_{MB}$ of the input ionizing continuum is
left to escape from the back of the clouds.  This fraction was
initially set to $\sim 60$\% (the value used in Ferruit
et\,al. 1997) but was allowed to increase substantially for the models
making use of the steep SED. This was necessary in order to keep a
high value of the
\ion{He}{2}/\hbeta\ ratio emitted by the MB component, an essential
ingredient of the dual-component models.  The ``filtered'' continuum
is later used to ionize the IB clouds further out (which do not emit
any \ion{He}{2}). The IB clouds integration is stopped when the
electron temperature falls below 5000\,K. Dust is considered to be
mixed with the ionized gas, and the code includes heating by dust
photoionization.  The dust content of the photoionized plasma is
described by the quantity $\mu$ which is the dust-to-gas ratio of the
plasma expressed in units of the solar neighborhood dust-to-gas ratio
($\mu=1$). In this work we use a constant $\mu$ = 0.015, which has a
negligible effect on line transfer and on the gas heating.

\begin{figure}
\plotone{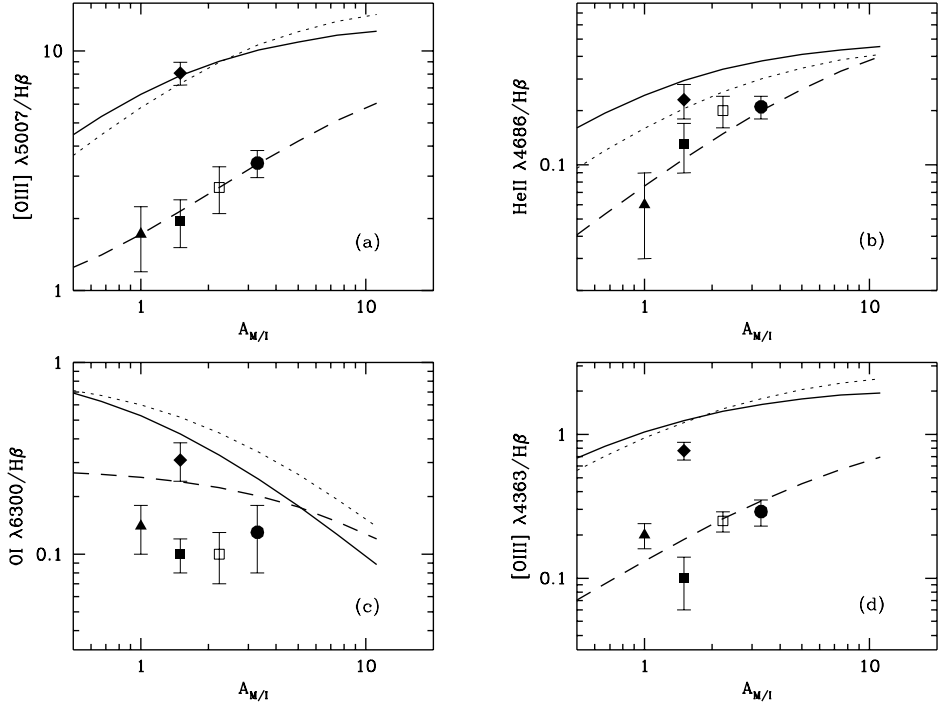}
\caption{Theoretical emission line ratios as function of the relative
proportion of MB and IB clouds, $A_{M/I}$. The solid line corresponds to the
predicted narrow line ratios under SED\,1 and the long-dashed line to
the output of SED\,2 (see text for further details). The short-dashed
line shows the effects on the line ratios by the inclusion of UV
absorbing material (WA with $U_{WA}$=0.01) between the BLR and
NLR. The filled triangle, square and circle and open square are the
observed dereddened line ratios of the NLS1 galaxies CTS\,H34.06,
1H\,1934-093 CTS\,R12.02 and MRK\,1239, respectively. The diamond corresponds to the
normal Seyfert 1 galaxy NCG\,5548. 
\label{diagramas}}
\end{figure}

Figure~\ref{diagramas} shows the predicted [\ion{O}{3}] \lb5007,
\ion{He}{2} \lb4686, [\ion{O}{1}] \lb6300 and [\ion{O}{3}] \lb4363, 
relative to \hbeta\ as function of the relative proportion of MB and
IB clouds, A$_{\rm M/I}$. The solid line corresponds to the
predictions of SED\,1 and the dashed line to the ratios obtained with
SED\,2.  In order to compare the output with the observations we have
plotted the dereddened observed ratios for the NLS1 galaxies
CTS\,H34.06 (filled triangle), 1H\,1934-063 (filled square),
CTS\,R12.02 (filled circle) and MRK\,1239 (open square) as taken
from Paper\,I. NGC\,5548 was chosen as representative of the normal
Seyfert 1 galaxies. Its dereddened ratios were taken from Kraemer et
al. 1998 and are shown as filled diamonds. Table 2
lists the predictions of the best-fit model for the most important
optical lines observed in each galaxy and the parameters of the
corresponding model.

A close inspection to Figure~\ref{diagramas} and the values of
Table 2 allows us to say that our approach is successful at
reproducing the observed ratios overall. The fact of being able to
predict simultaneously and accurately the [\ion{O}{3}] \lb5007/\hbeta\
and \ion{He}{2} \lb4686/\hbeta\ ratios for the NLS1 galaxies and
NCG\,5548 indicates that the general distribution of ionizing photons
at least in the range 13.6 -- 100 eV is correct.  Since the model
parameters were rather similar for the two SEDs, the results provide a
strong support to the idea that the NLR of NLS1s and Sy1s have similar
physical properties in terms of density, chemical abundance,
ionization parameter and distance to the central source.  The most
probable cause of the observed spectral differences can be related to
the variation in steepness of their ionizing continuum.

The largest discrepancies between models and observations, for some of
the galaxies, are in low excitation lines ratios such as
[\ion{O}{1}] \lb6300/\hbeta, [\ion{S}{2}] \lb\lb6717,6731/\hbeta\ and
[\ion{N}{2}] \lb6584/\hbeta.  These are overpredicted by a factor of 2
in CTS\,H34.06, 1H\,1034-063 and MRK\,1239. 

This systematic overprediction, nonetheless, might provide helpful
constraints to the physical conditions of the IB clouds and the nature
of the central ionizing source. In effect, it is well known that the
low ionization lines of AGNs are produced by the flux of high energy
EUV photons in the range 300--900 eV (cf fig~1 in Binette et~al. 1997)
which penetrate deep into the NLR gas. If these photons are not
allowed to reach this region, intrinsically weaker low ionization
lines will be emitted. A possible candidate for this shielding might
plausibly be a high ionization warm absorber (hereafter WA), located
near the BLR.  Evidences for the existence of such an absorbing
material have been found in several NLS1s, particularly in those with
high polarization (Leighly 1997). MRK\,1239, the NLS1 with the largest
discrepancies between the predicted and observed low ionization lines,
(cf.  Table 2) is widely known for being highly polarized
(Goodrich 1989) and for showing WA spectral features (Leighly et\,al. 
1997; Grupe et\, al. 1998). In contrast, CTS\,R12.02 (= NGC\,4748) 
shows one of the lowest percentage of polarization (0.12\% against 
2.89\% of MRK1239) according to Goodrich (1989) and presents here the 
best agreement between the model and the observed line ratios, even in 
the low ionization lines. No report about X-rays WA features are found 
in the literature for this galaxy.

Although polarization data for the remaining NLS1s are not available at 
present, the case of MRK\,1239 can be taken as a good evidence for 
atenuation of the hard X-ray continuum and its influence on the NLR 
emitted spectrum. We discard the possibility of higher densities for the 
IB clouds since the derived value using the [\ion{S}{2}] \lb6717/\lb6731 
ratio in the galaxies are near $10^{3}$ \cm.

Although a finer tuning of the parameters might improve the
agreement between the observations and the model predictions for the
objects studied here, this is unwarranted considering the
simplicity of the approach taken. What is important to note is that
the same initial set of physical conditions (albeit different SED)
derived independently for
normal Seyfert 1s are also valid to reproduce to a first
approximation the emission line ratios observed in NLS1s.

\section{New Emission Line Ratios for the NLR of NLS1s}\label{bla}

As was discussed in the Introduction and Section~\ref{nabc}, one of
the main purposes of this paper is to review the current assumption
that the NLR contributes \hbeta\ with 0.1$\times$ the flux of
[\ion{O}{3}] \lb5007, implying a universal constant [\ion{O}{3}]
\lb5007/\hbeta\ ratio of 10 (Osterbrock \& Pogge 1985, Leighly
1999). This assumption is usually made on account of the difficulty in
deblending the narrow and broad components of the permitted lines in
these objects and is based on trends observed in narrow emission lines
of Seyfert 2s and intermediate Seyfert 1 galaxies. According to the
results obtained in Section~\ref{gauss} the NLR of NLS1s
contributes 50\% of the total \hbeta\ emission. This has important
implications for the intrinsic ratios of the narrow lines. For
instance, [\ion{O}{3}] \lb5007/\hbeta\ (narrow) now falls in the
interval 1--5 since narrow \hbeta\ sees its flux increased by up
to ten times its previously assumed value.

If we had (erroneously) assumed that in NLS1s the
[\ion{O}{3}]\lb5007/\hbeta\ (narrow) ratio took the canonical value of 
10 of normal Seyferts, we would have inferred a spectral index $\le \alpha
\sim -1.4$ to describe the EUV-to-soft-X-ray SED  since such hardness
is favored when atempting to reproduce a high ratio. The ROSAT
observations, however, show that the soft X-ray spectra from NLS1s
is systematically steeper ($\alpha \sim -2$ or smaller) than those
from Seyfert 1 galaxies with broad optical lines. In addition,
Table 2 shows that photoionization models with $\alpha \sim
-2.2$ always provide [\ion{O}{3}]\lb5007/\hbeta\ ratios smaller
than 7 under the initial set of conditions assumed. This result is
therefore fully consistent with our findings of a much lower ratio in
NLSIs than in normal Seyferts.  Beside, much lower NLR densities
($n_{B} \sim 10^{5}$ \cm) would have been required to get [\ion{O}{3}]
\lb5007/\hbeta\ around 10 using SED\,2. Such low densities would be
discrepant with the higher ones determined in Paper~I and from other
authors (Osterbrock \& Pogge 1985).

One possibility is that the continuum in the 13.6--200 eV
region that photoionizes the NLR of NLS1s is similar to that of
NGC\,5548 (SED\,1), but that it is modified before reaching the NLR gas by
intervening material located between the BLR and NLR.  Kraemer
et~al. (1999) explored, for instance, the effects of UV absorbing
material on the shape of the continuum radiation emitted from the AGN,
and on the relative strengths of the ensuing emission lines formed in
the NLR of Seyfert 1 galaxies exposed to this emerging continuum 
distribution. Their results indicated that a low ionization UV absorber 
with a large covering factor can indeed modify the intrinsic EUV 
continuum of the central source and produce significative variations in 
the NLR emission spectrum of the AGN compared to that produced by the 
unattenuated continuum. Nonetheless, amongst the various types of warm 
absorbers tested, no model could reproduce a low [\ion{O}{3}] 
\lb5007/\hbeta\ ratio. It should be mentioned that they assumed the NLR 
gas to be uniformly radiation-bounded unlike the models presented here
which consist of a combination of MB and IB components.

We consider worthy to test whether the inclusion of a low-ionization
WA with similar characteristics to that of Kraemer et\,al. (1999)
and exposed to SED\,1 might explain the strongest (narrow) line
ratios observed in NLS1s. In this picture, differences in
[\ion{O}{3}] \lb5007/\hbeta\ for example would be caused by the
effects of the intervening WA rather than by changes in steepness
of the intrinsic EUV -- soft X-ray continuum, as was suggested in the
preceding section. Results are presented in
Fig.~\ref{diagramas} where the short-dashed line shows the predicted
line ratios for a NLR with the same physical parameters as
employed in modeling NCG\,5548 but photoionized by the continuum
leaking from a WA with $U_{WA}$=0.01, $n_{H}=1\times10^{7}$ \cm, solar
abundance and thickness N$_{\rm H}=10^{20}$ cm$^{-2}$. The
intrinsic SED incident on the WA was SED\,1. 

Comparison of these
results with those of the unabsorbed model (solid line) allows us
to say that the variations of the emission line ratios introduced
by an intervening low-ionization WA are relatively minor and clearly
insufficient for explaining the differences in line ratios observed
between NLS1s and normal Sy1s.  On the other hand, models with
$U_{WA} < 10^{-2.5}$ are found to alter the incident
EUV-to-soft-X-ray distribution in such a way that few photons are left
to ionize the NLR as in Kraemer et\,al. (1999). It can be seen that the
[\ion{O}{3}] \lb5007/\hbeta\ ratio does not change significantly
relative to the unabsorbed model. In either cases, low ionization
lines such as [\ion{O}{1}] \lb6300 (Figure~\ref{diagramas}c) see
their ratios increase relative to \hbeta. The same trend is observed
for [\ion{N}{2}] \lb6584 and [\ion{S}{2}] \lb\lb6717,6731. We
recall from the previous section that these lines were already
overpredicted.  The lower the ionization degree of the WA, the
larger this effect. This is clearly understood if we consider that the
net effect of a low-ionization WA is to modify the EUV
distribution while the hard EUV portion is left unaltered. Since
low ionization lines are produced by the harder EUV radiation,
they are now augmented due to the reduction of ionizing photons
which produce the Balmer and \ion{He}{2} \lb4686 lines.

We therefore conclude that a low-ionization WA, if it exists in
NLS1s, cannot explain the low [\ion{O}{3}] \lb5007/\hbeta\ ratio
observed in these objects.

\section{Conclusions}\label{fin}

We have analyzed long-slit spectral data of a sample composed of seven NLS1 
galaxies. A decomposition of the \halfa\ and \hbeta\ emission line profiles 
into Gaussian components allowed us to separate the flux contribution of the 
NLR from the total flux of the line. Our results show that, on average, 50\% 
of the total \hbeta\ flux is due to emission from the NLR. Using the 
[\ion{O}{3}] \lb5007 line profile as a template for the narrow lines in order 
to subtract this contribution in the permitted lines give very similar results 
to those obtain throught the Gaussian decomposition. This confirms the 
presence of a broad component in the permitted lines.

The FWHM of the broad components of \halfa\ and \hbeta\ in the NLS1s studied
here seems to be rather uniform within the same galaxy  after comparing \halfa\ and
\hbeta\  and throughout the
sample (2250 \kms\ and 2560 \kms\ for \halfa\ and \hbeta, respectively).
No evidence of Lorentzian profiles was observed neither in the narrow nor the
broad lines. The narrow components of  \halfa\ and \hbeta\ present
FWHM comparable to those of the forbidden lines which are typical of the NLR
in any Seyfert.

The resulting [\ion{O}{3}] \lb5007/\hbeta\ ratios fall in the interval
1--5, significantly lower than the value currently assumed
($\sim$10). This entails that the emission line ratios from the
NLR are different in NLS1s from those observed in normal and
intermediate Sy1 galaxies.

We test photoionization models that consider a NLR composed of a
combination of matter-bounded clouds and ionization-bounded
clouds. The former, with typical densities $\sim 10^{6}$ \cm\ and
photoionized by the intrinsic 
continuum from the central source, are responsible
for the emission of most of the [\ion{O}{3}] and high ionization lines.
This component should be located in the inner regions of the NLR. The 
latter, located farther out than the MB clouds and characterized by a 
lower density ($n_{e} \sim 10^{3}$ \cm), are photoionized by the 
continuum filtered from the MB clouds and emits most of the low 
ionization lines. Assuming similar physical parameters in the NLR of 
NLS1s and normal Sy1s, we show that the observed differences in 
emission line ratios between these two groups of galaxies can be 
explained in terms of differences in the form of the input ionizing 
spectra. NLS1s ratios are better reproduced with a steep power-law 
continuum, with spectral index $\alpha < -2$ while flatter spectral 
indices ($\alpha \sim -1.5$) match the observed line ratios in normal 
Sy1s. This scenario reproduces with very good agreement the line ratios
of NLS1s. It is furthermore consistent with ROSAT observations of
NLS1s, which show that these objects are characterized by steeper
power-law indices than those of Sy1 galaxies with broad optical
lines. Our modeling therefore support the view that the NLR is
directly photoionized by the unaltered SED distribution emitted by
the central engine.
  
\clearpage

\acknowledgments

ARA gratefully acknowledges the staff of the CASLEO Observatory for
instrumental and observing assistance. This research has made use of the 
NASA/IPAC Extragalactic Database (NED) which is operated by the Jet Propulsion 
Laboratory, Californian Institute of Technology, under contract with the 
National Aeronautics and Space Administration. The work of Luc Binette
was supported by the CONACyT grant 27546-E and the work of ARA and MGP
was supported by the PRONEX/FINEP grant  76.97.1003.00

\clearpage


\onecolumn

\begin{deluxetable}{lcccccccccc}
\scriptsize
\tablewidth{0pt}
\tablecaption{FWHM\tablenotemark{1} \,and flux line ratios in NLS1s. \label{fwhm}}
\tablehead{
\colhead{} & \multicolumn{2}{c}{FWHM H$\alpha$} & \colhead{} &  \multicolumn{2}{c}{FWHM H$\beta$} & 
\colhead{} & \colhead{FWHM} & \colhead{FWHM} & \colhead{} & \colhead{} \\
\cline{2-3}  \cline{5-6}
\colhead{Galaxy} & \colhead{Narrow} & \colhead{Broad} & \colhead{} & \colhead{Narrow} & 
\colhead{Broad} & \colhead{} & \colhead{$\lambda$5007} & \colhead{$\lambda$6584} & 
\colhead{F(\hbeta$_{\rm n}$/\hbeta$_{\rm b}$)} &
\colhead{F($\lambda$5007/\hbeta$_{\rm n}$)} \\
~~~(1) & (2) & (3) &  & (4) & (5) & & (6) & (7) & (8) & (9)}
\startdata 
1H\,1934-063   &     743  & 2703 &  & 926   & 2706 &  & 562   & 446  & 1.6  & 2.2 \nl
CTS\,H34.06    &     552  & 2401 &  & 972   & 2109 &  & 560   & 420  & 1.2  & 1.8 \nl
CTS\,J03.19    &     360  & 1766 &  & 560   & 1932 &  & 570   & 360  & 0.5  & 5.0 \nl
CTS\,J04.08    &     420  & 2152 &  & 560   & 2628 &  & 584   & 420  & 0.5  & 3.3 \nl
CTS\,J13.12    &     635  & 2212 &  & 762   & 3155 &  & 590   & 360  & 2.2  & 0.8 \nl
MRK\,1239      &     637  & 2278 &  & 683   & 2968 &  & 565   & 360  & 1.1  & 2.7 \nl
CTS\,R12.02    &     625  & 2272 &  & 1200  & 2434 &  & 583   & 448  & 0.7  & 3.4 \nl
\enddata
\tablenotetext{1}{In units of \kms}
\end{deluxetable}

\clearpage

\begin{deluxetable}{lcccccccccc}
\scriptsize
\tablewidth{0pt}
\tablecaption{Line Ratios (Relative to \hbeta) from Models and Observations. \label{models}}
\tablehead{
\colhead{} & \multicolumn{2}{c}{CTS\,H34.06} & \multicolumn{2}{c}{1H\,1934-063} & 
\multicolumn{2}{c}{CTS\,R12.02} & \multicolumn{2}{c}{MRK\,1239} & \multicolumn{2}{c}{NGC\,5548} \\
\colhead{Line} & \colhead{Model} & \colhead{Obs.} & \colhead{Model} & \colhead{Obs.} &
\colhead{Model} & \colhead{Obs.} & \colhead{Model} & \colhead{Obs.} &
\colhead{Model} & \colhead{Obs.}}
\startdata
[\ion{Ne}{3}] \lb3869 & 0.36& 0.21$\pm$0.07& 0.44& 0.27$\pm$0.07& 0.67&
0.45$\pm$0.07& 0.54& 0.18$\pm$0.02& 1.56& 1.32$\pm$0.18\nl 
[\ion{O}{3}] \lb4363  & 0.13& 0.20$\pm$0.04& 0.18& 0.10$\pm$0.04& 
0.34& 0.29$\pm$0.06& 0.26& 0.25$\pm$0.04& 1.25& 0.77$\pm$0.11\nl
\ion{He}{2} \lb4686   & 0.08& 0.06$\pm$0.03& 0.11& 0.13$\pm$0.04& 
0.20& 0.21$\pm$0.03& 0.15& 0.20$\pm$0.04& 0.29& 0.23$\pm$0.05\nl
[\ion{O}{3}] \lb5007  & 1.72& 1.79$\pm$0.52& 2.14& 1.95$\pm$0.40& 
3.40& 3.40$\pm$0.44& 2.69& 2.69$\pm$0.60& 7.87& 8.07$\pm$0.88\nl
\ion{He}{1} \lb5875   & 0.14& 0.15$\pm$0.02& 0.12& 0.22$\pm$0.05& 
0.10& 0.15$\pm$0.06& 0.13& 0.13$\pm$0.02& 0.11& 0.21$\pm$0.05\nl
[\ion{O}{1}] \lb6300  & 0.25& 0.14$\pm$0.04& 0.24& 0.10$\pm$0.02& 
0.20& 0.13$\pm$0.06& 0.22& 0.10$\pm$0.03& 0.42& 0.31$\pm$0.07\nl
[\ion{N}{2}] \lb6584  & 2.39& 0.72$\pm$0.10& 2.27& 0.69$\pm$0.13& 
1.91& 1.75$\pm$0.18& 2.11& 0.28$\pm$0.03& 1.82& 0.76$\pm$0.15\nl
[\ion{S}{2}] \lb6725\tablenotemark{1}&0.95&0.47$\pm$0.12&0.90&
0.33$\pm$0.08&0.75&0.72$\pm$0.06&0.83&0.16$\pm$0.04&1.28&0.66$\pm$0.10\nl
[\ion{S}{3}] \lb9525  & 1.28& 0.20$\pm$0.08& 1.22& 0.50$\pm$0.10& 1.04& 
0.60$\pm$0.15& 1.14& 0.10$\pm$0.04&  0.75& 0.68$\pm$0.13\nl
$A_{M/I}$             & 1.0  &            & 1.5  &          & 2.2  & 
          & 3.3  &           &    1.5  &              \nl
$U_{MB}$              & 10$^{-1.4}$ &     & 10$^{-1.4}$ &   & 10$^{-1.4}$ &
    & 10$^{-1.4}$ &    &    10$^{-1.5}$ &       \nl
$n_{H,MB}$ (cm$^{-3}$) & 10$^{6}$   &     & 10$^{6}$    &   & 10$^{6}$    & 
   & 10$^{6}$    &    &    10$^{6}$    &       \nl
SED                   & (1)         &     & (2)         &   & (2)         &
    & (2)         &    &    (2)         &       \nl
F$_{MB}$ (\%)         & 85          &     & 85          &   & 85          &
    & 85          &    &    54          &       \nl 
\enddata
\tablenotetext{1}{Sum of fluxes of [\ion{S}{2}] \lb\lb6717,6731}
\end{deluxetable}

\end{document}